\documentclass[12pt]{article}
\usepackage{amsbsy}
\usepackage{amsfonts}
\usepackage{amsmath}
\usepackage{amssymb}
\usepackage{apacite}
\usepackage{setspace}
\usepackage{graphicx}
\usepackage{bm,multicol}
\usepackage[english]{babel}
\usepackage{natbib}
\usepackage[T1]{fontenc}
\usepackage[utf8]{inputenc}
\usepackage{authblk}
\usepackage{xcolor}

\newtheorem{lemma}{Lemma}

\newcommand{\newc}{\newcommand}
\newc{\N}{\mbox{N}}
\newc{\1}{\bf{1}}

\begin{document}

\title{Simple Bayesian Testing of Scientific Expectations in Linear Regression Models}
\author[1,2]{J. Mulder}
\author[3]{A. Olsson-Collentine}
\affil[1]{Department of Methodology and Statistics, Tilburg University, Warrandelaan 1, Tilburg, the Netherlands. Email: j.mulder3@tilburguniversity.edu}
\affil[2]{Jheronimus Academy of Data Science, Sint Janssingel 92, 5211 DA 's-Hertogenbosch, The Netherlands}
\affil[3]{Department of Methodology and Statistics, Tilburg University, Warrandelaan 1, Tilburg, the Netherlands. Email: j.a.e.olssoncollentine@tilburguniversity.edu}
\date{}                     
\setcounter{Maxaffil}{0}
\renewcommand\Affilfont{\itshape\small}

\maketitle

\begin{abstract}
Scientific theories can often be formulated using equality and order constraints on the relative effects in a linear regression model. For example, it may be expected that the effect of the first predictor is larger than the effect of the second predictor, and the second predictor is expected to be larger than the third predictor. The goal is then to test such expectations against competing scientific expectations or theories. In this paper a simple default Bayes factor test is proposed for testing multiple hypotheses with equality and order constraints on the effects of interest.  
The proposed testing criterion can be computed without requiring external prior information about the expected effects before observing the data. The method is implemented in R-package called `{\tt lmhyp}' which is freely downloadable and ready to use. The usability of the method and software is illustrated using empirical applications from the social and behavioral sciences.
\end{abstract}


\section{Introduction}
The linear regression model is the most widely used statistical method for assessing the relative effects of a given set of predictors on a continuous outcome variable. This assessment of the relative effects is an essential part when testing, fine-graining, and building scientific theories. For example, in work and organizational psychology the regression model has been used to better understand the effects of discrimination by coworkers and managers on workers' well-being \citep{Johnson:2012}; in sociology to assess the effects of the different dimensions of socioeconomic status on one's attitude towards immigrants \citep{Scheepers:2002}; and in experimental psychology to make inferences regarding the effects of gender when hiring employees (Carlson and Sinclair, 2017). Despite the extensive literature on statistical tools for linear regression analysis, methods for evaluating multiple hypotheses with equality and order constraints on the relative effects in a direct manner are still limited. This paper presents a Bayes factor testing procedure with accompanying software for testing such hypotheses with the goal to aid researchers in the development and evaluation of scientific theories.

As an example, let us consider the following linear regression model where a dependent variable is regressed on three predictor variables, say, $X_1$, $X_2$, and $X_3$:
\begin{eqnarray*}
y_{i} &=& \beta_0 +\beta_{1} X_{i,1} + \beta_{2} X_{i,2} + \beta_{3} X_{i,3} + \epsilon_i,
\end{eqnarray*}
where $y_{i}$ is the dependent variable of the $i$-th observation, $X_{i,k}$ denotes the $k$ predictor variable of the $i$-th observation, $\beta_{k}$ is the regression coefficient of the $k$-th predictor, for $k=1,\ldots,3$, $\beta_0$ is the intercept, and $\epsilon_i$ are independent normally distributed errors with unknown variance $\sigma^2$, for $i=1,\ldots,n$.

In exploratory studies the interest is typically whether each predictor has an effect on the dependent variable, and if there is evidence of a nonzero effect, we would be interested in whether the effect is positive or negative. In the proposed methodology such an exploratory analysis can be executed by simultaneously testing whether an effect is zero, positive, or negative. For the first predictor, the exploratory multiple hypothesis test would be formulated as
\begin{eqnarray}
\nonumber H_0&:&\beta_{1}=0\\
\label{test1}H_1&:&\beta_{1}>0\\
\nonumber H_2&:&\beta_{1}<0.
\end{eqnarray}
The proposed Bayes factor test will then provide a default quantification of the relative evidence in the data between these hypotheses.

In confirmatory studies, the interest is typically in testing specific hypotheses with equality and order constraints on the relative effects based on scientific expectations or psychological theories \citep{Hoijtink:2011}. Contrasting regression effects against each other using equality or order constraints can be more informative than interpreting them at certain benchmark values (e.g., standardized effects of .2, .5, and 1, are sometimes interpreted as `small', `medium', and `large' effects, respectively) because effects are not absolute but relative quantifications; relative to each other and relative to the scientific field and context \citep{Cohen}. For example, a standardized effect of .4 may be important for an organizational psychologist who is interested in the effect of discrimination on well-being on the work floor but less so for a medical psychologist who wishes to predict the growth of a tumor of a patient through a cognitive test. As such, interpreting regression effects relative to each other using equality and order constraints would be more insightful than interpreting the effects using fixed benchmarks.

In the above regression model for instance, let us assume that $\beta_1$, $\beta_2$, and $\beta_3$ denote the effects of a strong, medium and mild treatment, respectively. It may then be hypothesized that the effect of the strong treatment is larger than the effect of the medium treatment, the effect of the medium treatment is expected to be larger than the effect of the mild treatment, and all effects are expected to be positive. Alternatively it may be expected that all treatments have an equal positive effect. These hypotheses can then be tested against a third hypothesis which complements the other hypotheses. This comes down to the following multiple hypothesis test:
\begin{eqnarray}
\nonumber H_1&:&\beta_{1} > \beta_{2}>\beta_{3}>0\\
\label{test2}H_2&:&\beta_{1} = \beta_{2}=\beta_{3}>0\\
\nonumber H_3&:&\text{neither $H_1$, nor $H_2$}.
\end{eqnarray}
Here the complement hypothesis $H_3$ covers the remaining possible values of $\beta_1$, $\beta_2$, and $\beta_3$ that do not satisfy the constraints under $H_1$ and $H_2$. Subsequently the interest is in quantifying the relative evidence in the data for these hypotheses.

A general advantage of Bayes factors for testing statistical hypotheses is that we obtain a direct quantification of the evidence in the data in favor of one hypothesis against another hypothesis. Furthermore, Bayes factors can be translated to the posterior probabilities of the hypotheses given the observed the data and the hypotheses of interest. These probabilities give a direct answer to the research question which hypothesis is most likely to be true and to what degree given the data. These posterior probabilities can be used to obtain conditional error probabilities of drawing an incorrect conclusion when `selecting' a hypothesis in light of the observed data. These and other properties have greatly contributed to the increasing popularity of Bayes factors for testing hypotheses in psychological research \citep{MulderWagenmakers:2016}.

The proposed testing criterion is based on the prior adjusted default Bayes factor \citep{Mulder:2014b}. The method has various attractive properties. First, the proposed Bayes factor has an analytic expression when testing hypotheses with equality and order constraints in a regression model. Thus computationally demanding numerical approximations can be avoided resulting in a fast and simple test. Furthermore, by allowing users to formulate hypotheses with equality as well as ordinal constraints a broad class of hypotheses can be tested in an easy and direct manner. Another useful property is that no proper (subjective) prior distribution needs to be formulated based on external prior knowledge, and therefore the method can be applied in an automatic fashion. This is achieved by adopting a fractional Bayes methodology \citep{OHagan:1995} where a default prior is implicitly constructed using a minimal fraction of the information in the observed data and the remaining (maximal) fraction is used for hypothesis testing \citep{Gilks:1995}. This default prior is then relocated to the boundary of the constrained space of the hypotheses. In the confirmatory test about the unconstrained default prior for $(\beta_1,\beta_3,\beta_3)$ would be centered around \textbf{0}. Because this Bayes factor can be computed without requiring external prior knowledge it is called a `default Bayes factor'. Thereby, these default Bayes factors differ from regular Bayes factors where a proper prior is specified reflecting the anticipated effects based on external prior knowledge \citep[e.g.,][]{Rouder:2015}. Other default Bayes factors that have been proposed in the literature are the fractional Bayes factor \citep{OHagan:1995}, the intrinsic Bayes factor \cite{Berger:1996}, and the Bayes factor based on expected-posterior priors \citep{PerezBerger:2002,Mulder:2009}.

Although various alternative testing procedures are available for hypothesis testing for linear regression analysis, these methods are limited to some degree. First, classical significance tests are only suitable for testing a null hypothesis against a single alternative, and unsuitable for testing multiple hypotheses with equality as well as order constraints \citep{Silvapulle:2004}. Second, traditional model comparison tools (e.g., the AIC, BIC, or CFI) are generally not suitable for evaluating models (or hypotheses) with order constraints on certain parameters \citep{Mulder:2009,Braeken:2015}. Third, currently available Bayes factor tests cannot be used for testing order hypotheses \citep{Rouder:2015}, are not computationally efficient \citep{Mulder:2012,Kluytmans:2012}, or are based on large sample approximations \citep{Gu:2017}. The proposed Bayes factor, on the other hand, can be used for testing hypotheses with equality and/or order constraints, is very fast to compute due to its analytic expression, and is an accurate default quantification of the evidence in the data in the case of small to moderate samples because it does not rely on large sample approximations. Other important properties of the proposed methodology are its large sample consistent behavior and its information consistent behavior \citep{Mulder:2014b,BoingMessing:2018}.

The Bayesian test is implemented in the R-package `{\tt lmhyp}', which is freely downloadable and ready for use in {\tt R}. The main function `{\tt test\_hyp}' needs a fitted modeling object using the `{\tt lm}' function together with a string that formulates a set of hypotheses with equality and order constraints on the regression coefficients of interest. The function computes the Bayes factors of interest as well as the posterior probabilities that each hypothesis is true after observing the data.

The paper is organized as follows. Section 2 presents the derivation of the default Bayes factor between hypotheses with equality and order hypotheses on the relative effects in a linear regression model. Section 3 presents the `{\tt lmhyp}' package and explains how it can be used for testing scientific expectations in psychological research. Section 4 shows how to apply the new procedure and software for testing scientific expectations in work and organizational psychology and social psychology. The paper ends with a short discussion.

\section{A default Bayes factor for equality and order hypotheses in a linear regression model}
\subsection{Model and hypothesis formulation}
For a linear regression model,
\begin{equation}
\textbf{y}=\textbf{X}\bm\beta+N(\textbf{0},\sigma^2\textbf{I}_n),
\label{model1}
\end{equation}
where $\textbf{y}$ is a vector of length $n$ of outcome variables, $\textbf{X}$ is a $n\times k$ matrix with the predictor variables, and $\bm\beta$ is a vector of length $k$ containing the regression coefficients, consider a hypothesis with equality and inequality constraints on certain regression coefficients of the form
\begin{equation}
H_t:\textbf{R}_E\bm\beta=\textbf{r}_E ~~~\&~~~ \textbf{R}_I\bm\beta>\textbf{r}_I,
\label{Ht}
\end{equation}
where $[\textbf{R}_E | \textbf{r}_E]$ and $[\textbf{R}_I|\textbf{r}_I]$ are the augmented matrices with $q_E$ and $q_I$ rows that contain the coefficients of the equality and inequality constraints, respectively, and $k+1$ columns. For example, for the regression model from the introduction, with $\bm\beta=(\beta_0,\beta_{1},\beta_{2},\beta_{3})'$, and the hypothesis $H_1:\beta_{1} > \beta_{2}>\beta_{3}>0$ in \eqref{test2}, the augmented matrix of the inequalities is given by
\[
[\textbf{R}_I|\textbf{r}_I] = \left[
\begin{array}{cccc|c}
0 & 1 & -1 & 0 & 0\\
0 & 0 & 1 & -1 & 0\\
0 & 0 & 0 & 1  & 0
\end{array}
\right]
\]
and for the hypothesis $H_2:\beta_{1} = \beta_{2}=\beta_{3}>0$, the augmented matrices are given by
\begin{eqnarray*}
[\textbf{R}_E|\textbf{r}_E] &=& \left[
\begin{array}{cccc|c}
0 & 1 & -1 & 0 & 0\\
0 & 0 & 1 & -1 & 0\\
\end{array}
\right]\\
\left[ \textbf{R}_I|\textbf{r}_I\right]  &=&
\left[
\begin{array}{cccc|c}
0 & 0 & 0 & 1 & 0
\end{array}
\right]
\end{eqnarray*}

The prior adjusted default Bayes factor will be derived for a constrained hypothesis in \eqref{Ht} against an unconstrained alternative hypothesis, denoted by $H_u:\bm\beta\in\mathbb{R}^k$, with no constraints on the regression coefficients. First we transform the regression coefficients as follows
\begin{equation}
\bm\xi=\left[\begin{array}{c}
\bm\xi_E\\
\bm\xi_I\end{array}\right]=\left[\begin{array}{c}
\textbf{R}_E\\
\textbf{D}\end{array}\right]\bm\beta=\textbf{T}\bm\beta,
\label{repara}
\end{equation}
where $\textbf{D}$ is a $(k-q_E)\times k$ matrix consisting of the unique rows of $\textbf{I}_k-\textbf{R}_E'(\textbf{R}_E\textbf{R}_E)^{-1}\textbf{R}_E$. 
Thus, $\bm\xi_E$ is a vector of length $q_E$ and $\bm\xi_I$ is a vector of length $k-q_E$. Consequently, model \eqref{model1} can be written as
\[
\textbf{y} = \textbf{X}\textbf{R}_E^{-1}\bm\xi_E + \textbf{X}\textbf{D}^{-1}\bm\xi_I + N(\textbf{0},\sigma^2\textbf{I}_n),
\]
because
\[
\textbf{X}\bm\beta=\textbf{X}\textbf{T}^{-1}\bm\xi=\textbf{X}\left[\textbf{R}_E^{-1}~\textbf{D}^{-1}\right]\left[\begin{array}{c}\bm\xi_E\\ \bm\xi_I\end{array}\right]=\textbf{X}\textbf{R}_E^{-1}\bm\xi_E+\textbf{X}\textbf{D}^{-1}\bm\xi_I,
\]
where $\textbf{R}_E^{-1}$ and $\textbf{D}^{-1}$ are the (Moore-Penrose) generalized inverse matrices of $\textbf{R}_E$ and $\textbf{D}$,
and the hypothesis in \eqref{Ht} can be written as
\begin{equation}
H_t:\bm\xi_E=\textbf{r}_E ~~~\&~~~ \tilde{\textbf{R}}_I\bm\xi_I>\tilde{\textbf{r}}_I,
\label{Ht2}
\end{equation}
because
\[
\textbf{R}_E\bm\beta=\textbf{R}_I\textbf{T}^{-1}\bm\xi=
\textbf{R}_E\left[\textbf{R}_E^{-1}~\textbf{D}^{-1}\right]\bm\xi=\left[\textbf{I}_{q_E}~\textbf{0}\right]\bm\xi=\bm\xi_E=\textbf{r}_E
\]
and
\begin{eqnarray*}
&&\textbf{R}_I\bm\beta=\textbf{R}_I\textbf{T}^{-1}\bm\xi=\textbf{R}_I\left[\textbf{R}_E^{-1}~\textbf{D}^{-1}\right]
\left[\begin{array}{c}
\textbf{r}_E\\
\bm\xi_I\end{array}\right]=\textbf{R}_I\textbf{R}_E^{-1}\textbf{r}_E+\textbf{R}_I\textbf{D}^{-1}\bm\xi_I>\textbf{r}_I\\
&&\Leftrightarrow \tilde{\textbf{R}}_I\bm\xi_I>\tilde{\textbf{r}}_I,~\text{with}~\tilde{\textbf{r}}_I=\textbf{r}_I-\textbf{R}_I\textbf{R}_E^{-1}\textbf{r}_E~\text{and}~\tilde{\textbf{R}}_I=\textbf{R}_I\textbf{D}^{-1}.
\end{eqnarray*}

\subsection{A default Bayes factor for testing hypotheses}
The Bayes factor for hypothesis $H_1$ against $H_2$ is defined as the ratio of their respective marginal likelihoods,
\[
B_{12}=\frac{p_1(\textbf{y})}{p_2(\textbf{y})}.
\]
The marginal likelihood quantifies the probability of the observed data under a hypothesis \citep{Jeffreys,Kass:1995}. For example, if $B_{12} = 10$ this implies that the data were 10 times more likely to have been observed under $H_1$ than under $H_2$. Therefore, the Bayes factor can be seen as a relative measure of evidence in the data between two hypotheses. The marginal likelihood under a constrained hypothesis $H_t$ in \eqref{Ht} is obtained by integrating the likelihood over the order constrained subspace of the free parameters weighted with the prior distribution,
\begin{equation}
p_t(\textbf{y}) = \iint_{\textbf{R}_I\bm\beta>\textbf{r}_I} p_t(\textbf{y}|\bm\beta,\sigma^2)\pi_t(\bm\beta,\sigma^2)d\bm\beta d\sigma^2,
\label{marglike1}
\end{equation}
where $p_t(\textbf{y}|\bm\beta,\sigma^2)$ denotes the likelihood of the data under hypothesis $H_t$ given the unknown model parameters, and $\pi_t$ denotes the prior distribution of the free parameters under $H_t$. The prior quantifies the plausibility of possible values that the model parameters can attain before observing the data.

Unlike in Bayesian estimation, the choice of the prior can have a large influence on the outcome of the Bayes factor. For this reason ad hoc or arbitrary prior specification should be avoided when testing hypotheses using the Bayes factor. However, specifying a prior that accurately reflects one's uncertainty about the model parameters before observing the data can be a time-consuming and difficult task \citep{Berger:2006a}. A complicating factor in the case of testing multiple, say, 3 or more, hypotheses, is that priors need to be carefully formulated for the free parameters under all hypotheses separately. Because noninformative improper priors also cannot be used when computing marginal likelihoods, there has been an increasing interest in the development of default Bayes factors where ad hoc or subjective prior specification is avoided. In these default Bayes factors a proper default prior is often (implicitly) constructed using a small part of the data while the remaining part is used for hypothesis testing. An example is the fractional Bayes factor \citep{OHagan:1995} where the marginal likelihood is defined by
\begin{equation}
p_t(\textbf{y}) = \iint_{\textbf{R}_I\bm\beta>\textbf{r}_I} p_t(\textbf{y}|\bm\beta,\sigma^2)^{1-b}\pi_t(\bm\beta,\sigma^2|\textbf{y}^b)d\bm\beta d\sigma^2,
\label{marglike2}
\end{equation}
where the (subjective) proper prior in \eqref{marglike1} is replaced by a proper default prior based on a (minimal) fraction ``$b$'' of the observed data\footnote{The proper default prior in \eqref{marglike2} is obtained by updating the noninformative improper (independence) Jeffreys' prior, $\pi^N(\bm\beta,\sigma^2)\propto\sigma^{-2}$, with a fraction $b$ of the data: $\pi_t(\bm\beta,\sigma^2|\textbf{y}^b)\propto \pi^N(\bm\beta,\sigma^2)f_t(\textbf{y}|\bm\beta,\sigma^2)^{b}$ \citep{Gilks:1995}.}, and the likelihood is raised to a power equal to the remaining fraction ``$1-b$'', which is used for hypothesis testing.

In this paper an adjustment of fractional Bayes factor is considered where the default prior is centered on the boundary (or null value) of the constrained space. The motivation for this adjustment is two-fold. First when testing a precise hypothesis, say, $H_0:\beta=0$ versus $H_a:\beta\not=0$, Jeffreys argued that a default prior for $\beta$ under $H_1$ should be concentrated around the null value because if the null would be false, the true effect would likely to be close to the null, otherwise there would be no point in testing $H_0$. Second, when testing hypotheses with inequality or order constraints, the prior probability that the constraints hold serves as a measure of the relative complexity (or size) of the constrained space under a hypothesis \citep{Mulder:2010}. This quantification of relative complexity of a hypothesis is important because the Bayes factor balances fit and complexity as an Occam's razor. This implies that simpler hypotheses (i.e., hypotheses having ``smaller'' parameter spaces) would be preferred over more complex hypotheses in the case of an approximately equal fit. Only when centering the prior at 0 when testing $H_1:\beta<0$ versus $H_2:\beta>0$, both hypotheses would be considered as equally complex with prior probabilities of .5 corresponding to half of the complete parameter space of $\beta$ of all real values ($\mathbb{R}$).

Given the above considerations, the fractional Bayes factor is adjusted such that the default prior is (i) centered on the boundary of the constrained parameter space and (ii) contains minimal information by specifying a minimal fraction. Because the model consists of $k+1$ unknown parameters ($k$ regression coefficients and an unknown error variance), a default prior is obtained using a minimal fraction\footnote{Updating the noninformative Jeffreys prior $\pi^N(\bm\beta,\sigma)\propto\sigma^{-2}$ with a sample of $k+1$ observations yields a proper marginal distribution for $\bm\beta$ having a multivariate Student $t$ distribution with 1 degree of freedom, which is equivalent to a multivariate Cauchy distribution.} of $b=\frac{k+1}{n}$.

In order to satisfy the prior property (i) when testing a hypothesis \eqref{Ht2}, the prior for $\bm\beta$ under the alternative should thus be centered at $\textbf{R}^{-1}\textbf{r}$, where $\textbf{R}'=[\textbf{R}_E'~\textbf{R}_I']$ and $\textbf{r}'=(\textbf{r}_E',\textbf{r}_I')$, which is equivalent to centering the prior for $\bm\xi$ at $\bm\mu^0=({\bm\mu_E^{0}}',{\bm\mu_I^{0}}')'=\textbf{TR}^{-1}\textbf{r}=(\textbf{r}'_E,\bm\mu_I^{0'})'$, with $\tilde{\textbf{R}}_I\bm\mu_I^0=\tilde{\textbf{r}}_I$. The following lemma gives the analytic expression of the default Bayes factor of a hypothesis with equality and order constraints on the regression coefficients versus an unconstrained alternative.

\begin{lemma}\label{lemma1}
The prior adjusted default Bayes factors for an equality-constrained hypothesis, $H_1:\textbf{R}_E\bm\beta=\textbf{r}_E$, an order-constrained hypothesis, $H_2:\textbf{R}_I\bm\beta>\textbf{r}_I$, and a hypothesis with equality and order constraints, $H_3:\textbf{R}_E\bm\beta=\textbf{r}_E$, $\textbf{R}_I\bm\beta>\textbf{r}_I$, against an unconstrained hypothesis $H_u:\bm\beta\in\mathbb{R}^k$ are given by
\begin{eqnarray}
B_{1u}&=& \frac{f^E_1}{c^E_1}=
\label{fcE}\frac{t(\textbf{r}_E;\textbf{R}_E\hat{\bm\beta},s^2(n-k)^{-1}\textbf{R}_E(\textbf{X}'\textbf{X})^{-1}\textbf{R}_E',n-k)}
{t(\textbf{r}_E;\textbf{r}_E,s^2\textbf{R}_E(\textbf{X}'\textbf{X})^{-1}\textbf{R}_E',1)},\\
B_{2u}&=& \frac{f^I_2}{c^I_2}=
\label{fcI}\frac{\text{Pr}(\textbf{R}_I\bm\beta>\textbf{r}_I|\textbf{y},H_u)}{\text{Pr}(\textbf{R}_I\bm\beta>\textbf{r}_I|\textbf{y}^b,H_u)},\\
B_{3u}&=&\frac{f^E_3}{c^E_3} \times \frac{f^{I|E}_3}{c^{I|E}_3}
\label{fcE2}=\frac{t(\textbf{r}_E;\textbf{R}_E\hat{\bm\beta},s^2(n-k)^{-1}\textbf{R}_E(\textbf{X}'\textbf{X})^{-1}\textbf{R}_E',n-k)}
{t(\textbf{r}_E;\textbf{r}_E,s^2\textbf{R}_E(\textbf{X}'\textbf{X})^{-1}\textbf{R}_E',1)}\\
&&\times \frac{\text{Pr}(\tilde{\textbf{R}}_I\bm\xi_I>\tilde{\textbf{r}}_I|\bm\xi_E=\textbf{r}_E,\textbf{y},H_u)}{\text{Pr}(\tilde{\textbf{R}}_I\bm\xi_I>\textbf{r}^*_I|\bm\xi_E=\hat{\bm\xi}_E,\textbf{y}^b,H_u)},
\label{fcI2}
\end{eqnarray}
where $\textbf{r}^*_I=\tilde{\textbf{R}}_I\hat{\bm\xi}_I$, $t(\bm\xi;\bm\mu,\textbf{S},\nu)$ denotes a Student $t$ density for $\bm\xi$ with location parameter $\bm\mu$, scale matrix $\textbf{S}$, and degrees of freedom $\nu$, $\hat{\bm\beta}=(\textbf{X}'\textbf{X})^{-1}\textbf{X}'\textbf{y}$ is the maximum likelihood estimate (MLE) of $\bm\beta$ and $s^2=(\textbf{y}-\textbf{X}\hat{\bm\beta})'(\textbf{y}-\textbf{X}\hat{\bm\beta})$ is the sums of squares, and the (conditional) distributions are given by
\begin{eqnarray*}
\pi(\bm\beta|\textbf{y},H_u)&=&t(\bm\beta;\hat{\bm\beta},s^2(\textbf{X}'\textbf{X})^{-1}/(n-k),n-k)\\
\pi(\bm\beta|\textbf{y}^b,H_u)&=&t(\bm\beta;\textbf{R}^{-1}_I\textbf{r}_I,s^2(\textbf{X}'\textbf{X})^{-1},1)\\
\pi(\bm\xi_I|\bm\xi_E=\textbf{r}_E,\textbf{y},H_u)&=&t(\bm\xi_I;\bm\mu_I^N,\textbf{S}_I^N,n-k)\\
\pi(\bm\xi_I|\bm\xi_E=\textbf{r}_E,\textbf{y}^b,H_u)&=&t(\bm\xi_I;\bm\mu_I^0,\textbf{S}_I^0,1)
\end{eqnarray*}
with
\begin{eqnarray*}
\bm\mu_I^N&=&\textbf{D}\hat{\bm\beta}+\textbf{D}(\textbf{X}'\textbf{X})^{-1}\textbf{R}_E'(\textbf{R}_E(\textbf{X}'\textbf{X})^{-1}\textbf{R}_E')^{-1}(\textbf{r}_E-\textbf{R}_E\hat{\bm\beta})\\
\textbf{S}_I^N&=&\left(1+s^{-2}(\textbf{r}_E-\textbf{R}_E\hat{\bm\beta})'(\textbf{R}_E(\textbf{X}'\textbf{X})^{-1}\textbf{R}_E')^{-1}(\textbf{r}_E-\textbf{R}_E\hat{\bm\beta})\right)(n-k+q_E)^{-1}s^2\\
&&(\textbf{D}(\textbf{X}'\textbf{X})^{-1}\textbf{D}'-\textbf{D}(\textbf{X}'\textbf{X})^{-1}\textbf{R}_E'(\textbf{R}_E(\textbf{X}'\textbf{X})^{-1}\textbf{R}_E')^{-1}\textbf{R}_E(\textbf{X}'\textbf{X})^{-1}\textbf{D}')\\
\textbf{S}_I^0&=&\tfrac{s^2}{1+q^E}(\textbf{D}(\textbf{X}'\textbf{X})^{-1}\textbf{D}'-\textbf{D}(\textbf{X}'\textbf{X})^{-1}\textbf{R}_E'(\textbf{R}_E(\textbf{X}'\textbf{X})^{-1}\textbf{R}_E')^{-1}\textbf{R}_E(\textbf{X}'\textbf{X})^{-1}\textbf{D}'),
\end{eqnarray*}
\end{lemma}
\noindent\textbf{Proof:} Appendix A.\\

Note that the factors in \eqref{fcE} and \eqref{fcI2} are multivariate Savage-Dickey density ratio's \citep{Dickey:1971,Wetzels:2010,Mulder:2010}. These ratio's have an analytic expression because the marginal posterior and default prior have multivariate Student $t$ distributions. In {\tt R} these can be computed using the {\tt dmvt} function in the {\tt mvtnorm}-package \citep{Genz:2016}. 

The ratio's of (conditional) probabilities in \eqref{fcI} and \eqref{fcI2} can also be computed in a straightforward manner. If $\tilde{\textbf{R}}_I$ is of full row-rank then the transformed parameter vector, say, $\bm\eta_I=\tilde{\textbf{R}}_I\bm\xi_I$ has a Student $t$ distribution so that $\text{Pr}(\tilde{\textbf{R}}_I\bm\xi_I>\tilde{\textbf{r}}_I|\bm\xi_E=\textbf{r}_E,\textbf{y},H_u)=\text{Pr}(\bm\eta_I>\tilde{\textbf{r}}_I|\bm\xi_E=\textbf{r}_E,\textbf{y},H_u)$ can be computed using the {\tt pmvt} function from the {\tt mvtnorm}-package \citep{Genz:2016}. If the rank of $\tilde{\textbf{R}}_I$ is lower than $q^I$, then the probability can be computed as the proportion of draws from an unconstrained Student $t$ distribution satisfying the order constraints.

The posterior quantities in the numerators reflect the relative fit of a constrained hypothesis, denoted by ``$f$'', relative to the unconstrained hypothesis: a larger posterior probability implies a good fit of the order constraints and a large posterior density at the null value indicates a good fit of a precise hypothesis. The prior quantities in the denominators reflect the relative complexity of a constrained hypothesis, denoted by ``$c$'', relative to the unconstrained hypothesis: a small prior probability implies a relatively small inequality constrained subspace, and thus a `simple' hypothesis, and a small prior density at the null value corresponds to a large spread (variance) of possible values under the unconstrained alternative implying the null hypothesis is relatively simple in comparison to the unconstrained hypothesis.

\begin{figure}
	\begin{center}
		\includegraphics[width=14cm,keepaspectratio=true]{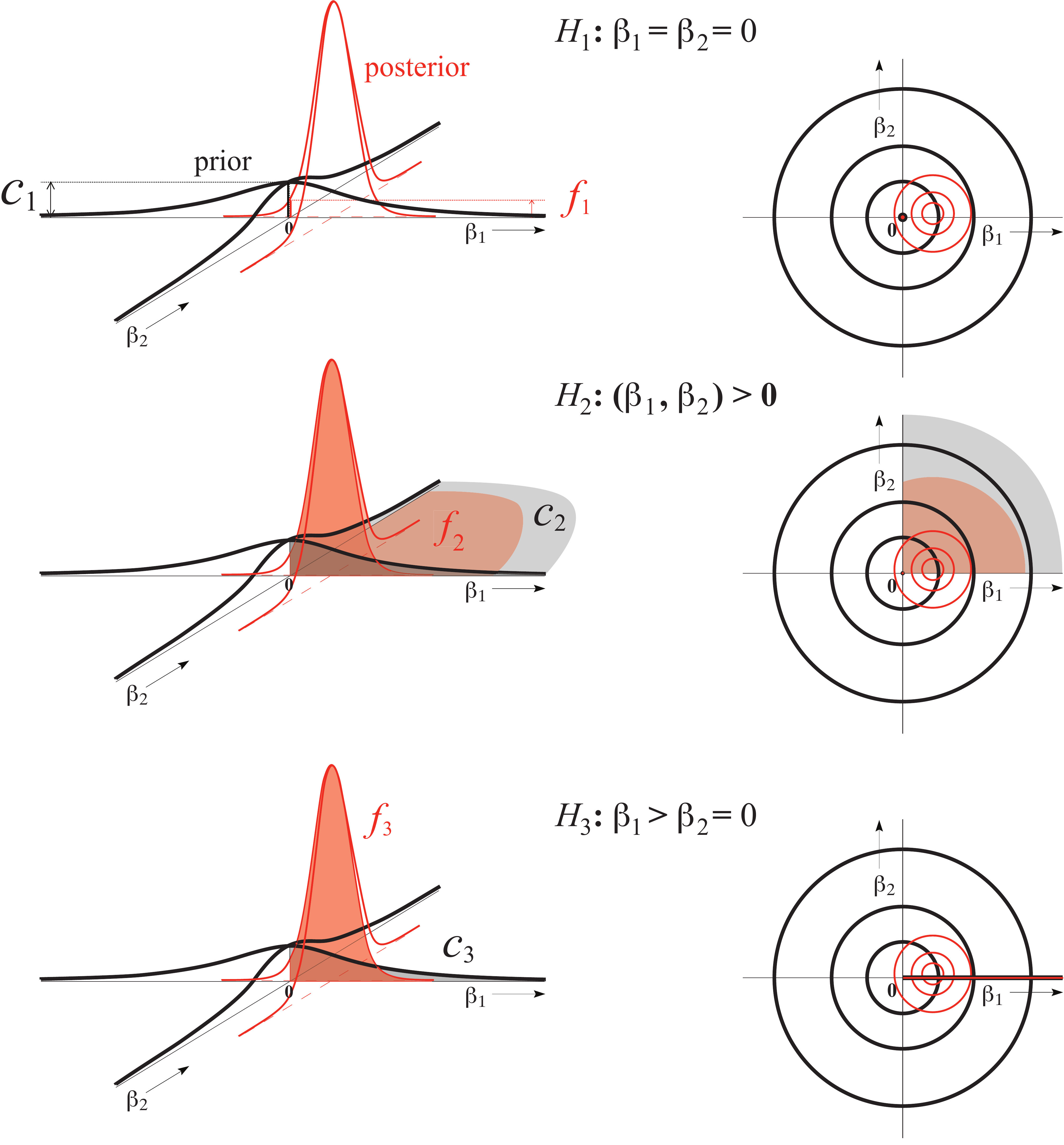}
	\end{center}
\caption{Graphical representation of the default Bayes factor for $H_1:\beta_1=\beta_2=0$ (upper panels), $H_2:\bm\beta>\textbf{0}$ (middle panels), and $H_3:\beta_1>\beta_2=0$ (lower panels) as the ratio's of the posterior (red, thin lines) and prior (black, thick lines) density at the null value, the posterior and prior probabilities, and the surfaces of the cross sections of the posterior and prior density, respectively.}
\label{fig_prior_post}
\end{figure}

Figure \ref{fig_prior_post} gives more insight about the nature of the expressions in \eqref{fcE} to \eqref{fcI2} in Lemma \ref{lemma1} for an equality constrained hypothesis, $H_1:\beta_1=\beta_2=0$ (upper panels), an inequality constrained hypothesis, $H_2:\bm\beta>\textbf{0}$ (middle panels), and hypothesis with an equality constraint and an inequality constraint, $H_3:\beta_1>\beta_2=0$ (lower panels). The Bayes factor for $H_1$ against the unconstrained hypothesis $H_u$ in \eqref{fcE} corresponds to the ratio of the unconstrained posterior density and the unconstrained default prior (which has a multivariate Cauchy distribution centered at the null value) evaluated at the null value. The Bayes factor for $H_2$ against $H_u$ in \eqref{fcI} corresponds to the ratio of posterior and default prior probabilities that the constraints hold under $H_u$. In the case of independent predictors, for example, the prior probability would be equal .25 as a result of centering the default prior at \textbf{0}. The inequality constrained hypothesis would then be quantified as 4 times less complex than the unconstrained hypothesis. Finally for a hypothesis with equality and inequality constraints, $H_3:\beta_1>\beta_2=0$, the Bayes factor in \eqref{fcE2}-\eqref{fcI2} corresponds to the ratio of the surfaces of cross section of the posterior and prior density on the line $\beta_1>0$, $\beta_2=0$.

The default Bayes factors between these hypotheses are computed for a simulated data set with MLEs $(\hat{\beta}_1,\hat{\beta}_2)=(.7,.03)$ (Appendix B) that results in  $B_{1u}=\frac{f^E_1}{c^E_1}=\frac{0.061}{0.159}=0.383$, $B_{2u}=\frac{f^I_2}{c^I_2}=\frac{0.546}{0.250}=2.183$, $B_{3u}=\frac{f^E_3}{c^E_3}\times\frac{f^{I|E}_3}{c^{I|E}_3}=\frac{1.608}{0.318}\times \frac{0.996}{0.500}=10.061$. As will be explained in the next section, it is recommendable to include the complement hypothesis in an analysis. The complement hypothesis covers the subspace of $\mathbb{R}^2$ that excludes the subspaces under $H_1$, $H_2$, and $H_3$. In this example the Bayes factor of the complement hypothesis against the unconstrained hypothesis equals $B_{cu}=\frac{1-f^I_2}{1-c^I_2}=\frac{0.454}{0.750}=0.606$.

After having obtained the default Bayes factor of each hypothesis against the unconstrained hypothesis, Bayes factors between the hypotheses of interest can be obtained through the transitivity property of the Bayes factor, e.g., $B_{31}=\frac{B_{3u}}{B_{1u}}=\frac{10.061}{.383}=26.299$. This implies there is strong evidence for $H_3$ relative to $H_1$, as the data were approximately 26 times more likely to have been produced under $H_3$ than under $H_1$.

%

Once the default Bayes factors of the hypotheses of interest against the unconstrained hypothesis are computed using Lemma 1, posterior probabilities can be computed for the hypotheses. In the case of, say, four hypotheses of interest against, the posterior probability that hypothesis $H_t$ is true can be obtained via
\begin{equation}
\text{Pr}(H_t|\textbf{y})= \frac{B_{tu}\text{Pr}(H_t)}{B_{1u}\text{Pr}(H_1)+B_{2u}\text{Pr}(H_2)+B_{3u}\text{Pr}(H_3)
+B_{cu}\text{Pr}(H_c)},
\label{PHP}
\end{equation}
for $t=1,$ 2, or 3, where $\text{Pr}(H_t)$ denotes the prior probability of hypothesis $H_t$, i.e., the probability that $H_t$ is true before observing the data. As can be seen, the posterior probability is a weighted average of the Bayes factors weighted with the prior probabilities. Throughout this paper we will work with equal prior probabilities, but other choices may be preferred in specific applications \citep[e.g.,][]{WagenmakersEtAl2011Bem}. For the example data from Appendix B and Figure \ref{fig_prior_post}, the posterior probabilities would be equal to $P(H_1|\textbf{y})=0.029$, $P(H_2|\textbf{y})=0.165$, $P(H_3|\textbf{y})=0.760$, and $P(H_c|\textbf{y})=0.046$. Based on these outcomes we would conclude that there is most evidence for $H_3$ that the effect of the first predictor is positive and the effect of the second predictor is zero with a posterior probability of .76. In order to draw a more decisive conclusion (e.g., when obtaining a posterior probability for a hypothesis larger than, say, .99) more data are needed.

\section{Software}
The Bayes factor testing criterion for evaluating inequality and order constrained hypotheses was implemented in a new {\tt R} package called `{\tt lmhyp}' to ensure general utilization of the methodology\footnote{Run `{\tt devtools::install\_github("jaeoc/lmhyp")}' in R to install the package.}. As input the main function `{\tt test\_hyp}' needs a fitted linear regression modeling object from the {\tt lm}-function as well as a string that specifies the constrained hypotheses of interest.

As output the function provides the default Bayes factors between all pairs of hypotheses. By default a complement hypothesis is also included in the analysis. For example, when testing the hypotheses, say, $H_1:\beta_1>\beta_2>\beta_3>0$ versus $H_2:\beta_1=\beta_2=\beta_3>0$, a third complement hypothesis $H_3$ will be automatically added which covers the remaining parameter space, i.e., $\mathbb{R}^3$ excluding the subspaces under $H_1$ and $H_2$. The reason for including the complement hypothesis is that Bayes factors provide relative measures of evidence between the hypotheses. For example, it may be that $H_2$ receives, say, 30 times more evidence than $H_1$, i.e., $B_{21}=30$, which could be seen as strong evidence for $H_2$ relative to $H_1$, yet it may be that $H_2$ still badly fits to the data in an absolute sense. In this case the evidence for the complement hypothesis $H_3$ against $H_2$ could be very large, say, $B_{32}=100$.

Besides the default Bayes factor the function also provides the posterior probabilities of the hypotheses. Posterior probabilities may be easier for users to interpret than Bayes factors because the posterior probabilities sum up to 1. Note that when setting equal prior probabilities between two hypotheses, the posterior odds of the hypotheses will be equal to the Bayes factor. By default all hypotheses receive equal prior probabilities. Thus, in the case of $T$ hypotheses, then $P(H_t)=\frac{1}{T}$, for $t=1,\ldots,T$. Users can manually specify the prior probabilities by using the `{\tt priorprobs}' argument. In the remaining part of the paper we will work with the default setting of equal prior probabilities. A step-by-step guide for using the software will be provided in the following section.

\section{Application of the new testing procedure using the software package `{\tt lmhyp}'}
In this section we illustrate how to use the `{\tt lmhyp}' package to test hypotheses, by applying the procedure to two empirical examples from psychology. We begin by describing the published research. In the two following subsections we then formulate hypotheses for each example and test these using the function \texttt{test\_hyp} from our R-package \texttt{lmhyp}.\footnote{The R-script used to produce the results in this section is available at https://osf.io/g8c9p/}

For the first example we use data from a study of mental health workers in England \citep{Johnson:2012}. The data of Johnson et al. measured health workers' well-being and its correlates, such as perceived discrimination from managers, coworkers, patients and visitors. Well-being was operationalized by scales measuring anxiety, depression and job dissatisfaction, the first two scales consisting of three items and the latter of five. The perceived discrimination variables are binary variables  that were meant to capture whether the worker believed they had been discriminated against from the four different sources in the last 12 months. This example demonstrates hypothesis testing in regards to single variables and the "exploratory" option of the {\tt test\_hyp} function.

Our second empirical example comes from research by \cite{Carlsson:2017}. Over four experiments, Carlson and Sinclair compare two theoretical explanations for perceptions of gender discrimination in hiring, although we use data from only the first experiment (available at https://osf.io/qcdgp/). In this study Carlson and Sinclair showed university students two fictive job applications from a man and a woman for a position as either a computer specialist or nurse. Participants were told that the fictive job applications had been sent to real companies as part of a previous study, but that only one of the two applicants had been invited to a job interview despite being equally qualified. A two-item scale was then used to measure participants' belief the outcome was due to gender discrimination. Several potential correlates were also measured using two-item scales, such as the individual's belief that (wo)men are generally discriminated against, their expectation that they are gender-stereotyped by others (`stigma consciousness') and the extent to which they identify as feminists. This example demonstrates testing hypotheses involving multiple variables.

\subsection{Hypothesis testing of single effects in organizational psychology}
In our first example we illustrate how our approach might be used to explore competing hypotheses for single variables. It is common when testing the effect of an independent variable in regression to look at whether it is significantly different from zero, or to do a one-sided test of a positive versus a negative effect. When using a Bayes factor test we can test all these hypotheses directly against each other and compare the relative evidence for each hypothesis.

\cite{Braeken:2015} theorized that work-place discrimination has a negative impact on workers' well-being. Here we are testing this expectation against a positive effect and a zero effect, while controlling for discrimination from different sources. For example, in the case of discrimination by managers we have
\begin{equation}
\begin{aligned}
H_1: \beta_{manager} < 0 \\
H_2: \beta_{manager} = 0 \\
H_3: \beta_{manager} > 0,
\end{aligned}
\end{equation}
while controlling for discrimination by coworkers, patients, and visitors through the following regression model
\begin{eqnarray*}
y_{anxiety,i} &=& \beta_0 + \beta_{manager}  X_{manager,i} + \beta_{coworkers}  X_{coworker,i}\\
&& + \beta_{patient}  X_{patient,i} + \beta_{visitor}  X_{visitor,i} + \text{error}_i
\end{eqnarray*}
where the $\beta$'s are the regression effects of the various sources of discrimination on anxiety.

Evaluating these three hypotheses in R is straightforward with the \texttt{test\_hyp} function from our R-package \texttt{lmhyp}. This function takes as arguments `{\tt object}', a fitted object using the {\tt lm} function, `\texttt{hyp}', a string vector specifying one or several hypotheses (separated by semicolons), `{\tt priorprob}', specifying the prior probabilities of each hypotheses (by default equal, {\tt priorprob = 1}),  and `{\tt mcrep}', an integer that specifies the number of draws to compute the prior and posterior probabilities in the (unusual) case the matrix with the coefficients of the order constraints is not of full row rank (by default {\tt mcrep=1e6}). In addition, the argument {\tt hyp} also allows as input the string "exploratory", which will test the likelihood of the data for a zero, positive, or negative effect of all variables in the regression model, including the intercept. We will make use of this functionality below, after first discussing how to test the three hypotheses for a single variable. To test the hypotheses, we first fit a linear model on the variables as usual:

\begin{verbatim}
fit <- lm(anx ~ discM + discC + discP + discV, data = dat1)
\end{verbatim}

Next,  hypotheses are specified in R as character strings using the variable names from the fitted linear model. It is possible to test the traditional null hypothesis of $\beta_{manager}=0$ against the two-sided alternative example $\beta_{manager}\not =0$ by writing

\begin{verbatim}
H2 <- "discM = 0"
\end{verbatim}

\noindent Note that the complement hypothesis, $\beta_{manager}\not =0$, is automatically included. However, by testing whether the effect is zero, positive, or negative simultaneously, we obtain a more complete picture of the possible existence and direction of the population effect. This can be achieved by specifying all hypotheses as a single character vector in which the hypotheses are separated by semicolons:

\begin{verbatim}
Hyp1v2v3 <- "discM < 0; discM = 0; discM > 0"
\end{verbatim}

\noindent Note that spacing does not matter. Once the hypotheses have been specified, they are tested by simply inputting them together with the fitted linear model object into the function \texttt{test\_hyp}:

\begin{verbatim}
result <- test_hyp(fit, Hyp1v2v3)
\end{verbatim}

\noindent This will compute the default Bayes factors from Lemma 1 between the hypotheses, as well as the posterior probabilities for the hypotheses. The posterior probabilities are printed as the primary output:

\begin{verbatim}
## Hypotheses:
## 
##   H1:   "discM<0"
##   H2:   "discM=0"
##   H3:   "discM>0"
## 
## Posterior probability of each hypothesis (rounded):
## 
##   H1:   0.000
##   H2:   0.000
##   H3:   1.000
\end{verbatim}

As can be seen the evidence is overwhelmingly in favor of a positive effect of discrimination from managers on anxiety amongst health workers.  In fact when concluding that $H_3:\beta_{manager}>0$ is true, we would have a conditional error probability of drawing the wrong conclusion of approximately zero. To perform this test for all regression effects one simply needs to set the second {\tt hyp} argument equal to {\tt "exploratory"}: 

\begin{verbatim}
result <- test_hyp(fit, "exploratory")
\end{verbatim}

\noindent This option assumes that each hypothesis is equally likely \textit{a priori}. In the current example we then get the following output:

\begin{verbatim}
## Hypotheses:
## 
##   H1:   "X < 0"
##   H2:   "X = 0"
##   H3:   "X > 0"
## 
## Posterior probabilities for each variable (rounded), 
## assuming equal prior probabilities:
## 
##                H1    H2    H3
##             X < 0 X = 0 X > 0
## (Intercept) 0.000 0.000 1.000
## discM       0.000 0.000 1.000
## discC       0.005 0.780 0.216
## discP       0.003 0.628 0.369
## discV       0.007 0.911 0.082
\end{verbatim}

\noindent The posterior probabilities for discrimination by managers are the same as when tested separately. In regards to the other variables, there seems to be positive evidence that there is no effect of discrimination by coworkers, patients, or visitors on anxiety. Note that the evidence for this is not as compelling as for the effect of discrimination by managers, as can be seen from the conditional error probabilities of .216, .369, and .082, respectively, which are quite large. Therefore more data are needed in order to draw more decisive conclusions. Note here that classical significance tests cannot be used for quantifying the evidence in the data in favor of the null; the classical test can only be used to falsify the null. When a null hypothesis cannot be rejected we are left in a state of ignorance because we cannot reject the null but also not claim there is evidence for the null \citep{Wagenmakers:2007}. 

Because the prior probabilities of the hypotheses are equal, the ratio of the posterior probabilities of two hypotheses corresponds with the Bayes factor, e.g., $B_{23}=\frac{\text{Pr}(H_2|\textbf{y})}{\text{Pr}(H_3|\textbf{y})}=\frac{.780}{.216}=3.615$, for the effect of discrimination by coworkers. By calling \texttt{BF\_matrix}, we obtain the default Bayes factors between all pairs of hypotheses. For convenience the printed Bayes factors are rounded to three digits, though exact values can be calculated from the posterior probabilities (unrounded posterior probabilities are available by calling \texttt{result\$post\_prob}). The Bayes factor matrix for {\tt discC} (discrimination from coworkers) can be obtained by calling

\begin{verbatim}
result$BF_matrix$discC
##         H1    H2    H3
## H1   1.000 0.006 0.022
## H2 162.367 1.000 3.615
## H3  44.913 0.277 1.000
\end{verbatim}

\noindent Hence, the null hypothesis of no effect is 162 times more likely than hypothesis $H_1$ which assumes a negative effect ($B_{21} = 162.367$), but only 3.6 times more likely than hypothesis $H_3$ which assumes a positive effect ($B_{23} = 3.615$). Similar Bayes factor matrices can be printed for all variables when using the "exploratory" option.

To summarize the first application, regressing the effects of perceived discrimination from managers, coworkers, patients, and visitors on the anxiety levels of English health workers, we found very strong evidence for a positive effect of perceived discrimination from managers on anxiety, mild to moderate evidence for no effect of discrimination from coworkers, patients, and visitors on anxiety. More research is needed to draw clearer conclusions regarding the existence of a zero or positive effect of these latter three variables.

\subsection{Hypothesis testing of multiple effects in social psychology}
In our second example we illustrate how our testing procedure can be used when testing multiple hypotheses with competing equal and order constraints on the effect of different predictor variables. Carlson and Sinclair (2017) compared two different theoretical explanations for perceptions of gender discrimination in hiring for the roles of computer specialist and nurse. To test individual differences they regressed perceptions of discrimination towards female victims on belief in discrimination against women, stigma consciousness and feminist identification, while controlling for gender and belief in discrimination against men. As a regression equation this can be expressed as

\begin{eqnarray*}
y_{discriminationW,i}& =& \beta_0 + \beta_{beliefW}  X_{beliefW,i} + \beta_{stigma}  X_{stigma,i} + \beta_{feminist}  X_{feminist} \\
&&+ \beta_{gender}  X_{gender,i} + \beta_{beliefM}  X_{beliefM,i} + \text{error}_i.
\end{eqnarray*}

\noindent where the $\beta$'s are standardized regression effects of the variables on perceived discrimination. Since in this subsection we will compare the beta-coefficients of different variables against each other, it facilitates interpretation if they are on the same scale. As such we standardize all variables before entering them in the model.

The two theories that Carlson and Sinclair (2008) examined make different explanations for what individual characteristics are most important to perceptions of gender discrimination. The `prototype explanation' suggests that what matters are the individual's beliefs that the gender in question is discriminated against, whereas the `same-gender bias explanation' suggests that identification with the victim is most important. In our example, the victim of discrimination is female and Carlson and Sinclair operationalize identification with the victim as stigma consciousness and feminist identity. Note that neither theory makes any predictions regarding the control variables (gender and general belief that men are discriminated against). A first hypothesis, based on the prototype explanation, might thus be that belief in discrimination of women in general is positively associated with the belief that the female applicant has been discriminated against, whereas stigma consciousness and feminist identity have no effect on this belief. Formally, this can be expressed as

\begin{equation}
H_1: \beta_{beliefW} > \beta_{stigma}= \beta_{feminist} = 0
\end{equation}

\noindent which is equivalent to:

\begin{equation}
H_1: \beta_{beliefW} > (\beta_{stigma}, \beta_{feminist}) = 0
\end{equation}

Alternatively, we might expect all three variables to have a positive effect on the dependent variable (all $\beta$'s > 0), but that, in accordance with the prototype explanation, a belief that women are generally discriminated against should have a larger effect on perceptions of discrimination than identifying with the job applicant. Formally this implies:

\begin{equation}
H_2: \beta_{beliefW} > (\beta_{stigma}, \beta_{feminist}) > 0
\end{equation}

A third hypothesis, based on the same-gender bias explanation, would be the reverse of the $H_1$, namely that stigma consciousness and feminist identity are positively associated with the outcome while a general belief in discrimination against women has no impact on the particular case. That is:

\begin{equation}
H_2:  (\beta_{stigma}, \beta_{feminist}) > \beta_{beliefW} = 0
\end{equation}

In this example we have thus specified three contradicting hypotheses regarding the relationships between three variables and wish to know which hypothesis receives most support from the data at hand. However, there is one additional implied hypothesis in this case: the complement. The complement, $H_c$, is the hypothesis that none of the specified hypotheses are true. The complement exists if the specified hypotheses are not exhaustive, that is, do not cover the entire parameter space. In other words, the complement exists if there are possible values for the regression coefficients which are not contained in the hypotheses, for example, $(\beta_1, \beta_2, \beta_3) = (-1, -1, -1)$ is a combination of effects which do not satisfy the constraints of either $H_1$, $H_2$, or $H_3$. Thus, the interest is in testing the following hypotheses:

\begin{eqnarray*}
H_1&:& \beta_{beliefW} > (\beta_{stigma}, \beta_{feminist}) = 0 \\
H_2&:& \beta_{beliefW} > (\beta_{stigma}, \beta_{feminist}) > 0  \\
H_3&:&  (\beta_{stigma}, \beta_{feminist}) > \beta_{beliefW} = 0 \\
H_c&:& \text{not }H_1, H_2, H_3
\end{eqnarray*}

As before, we begin by fitting a linear regression on the (standardized) variables:

\begin{verbatim}
fit <- lm(discW ~ beliefW + stigma + feminist + beliefM + gender,
          data = dat2)
\end{verbatim}

\noindent Next, we specify the hypotheses separated by semicolons as a character vector, here on separate lines for space reasons:

\begin{verbatim}
hyp1v2v3 <- "beliefW > (stigma, feminist) = 0; 
		  beliefW > (stigma, feminist) > 0;
		  (stigma, feminist) > beliefW = 0"
\end{verbatim}

\noindent The complement does not need to be specified, as the function will include it automatically if necessary. For this example we get the following output:

\begin{verbatim}
## Hypotheses:
## 
##   H1:   "beliefW>(stigma,feminist)=0"
##   H2:   "beliefW>(stigma,feminist)>0"
##   H3:   "(stigma,feminist)>beliefW=0"
##   Hc:   "Not H1-H3"
## 
## Posterior probability of each hypothesis (rounded):
## 
##   H1:   0.637
##   H2:   0.359
##   H3:   0.000
##   Hc:   0.004
\end{verbatim}

From the output posterior probabilities we see that $H_1$ and $H_2$, both based on the prototype explanation, received the most support, whereas $H_3$, which was derived from the same-gender bias model, and the complementary hypothesis are both highly unlikely. These results can be succinctly reported as: \textit{"Using a default Bayes factor approach, we obtain overwhelming evidence that either hypothesis $H_1$ or $H_2$ is true with posterior probabilities of approximately .637, .359, .000,  and .004 for $H_1$, $H_2$, $H_3$, and $H_4$, respectively."} Printing the Bayes Factor matrix yields:

\begin{verbatim}
result$BF_matrix
##       H1    H2       H3      Hc
## H1 1.000 1.776 1634.299 163.201
## H2 0.563 1.000  920.205  91.892
## H3 0.001 0.001    1.000   0.100
## Hc 0.006 0.011   10.014   1.000
\end{verbatim}

\noindent We see that the evidence for both $H_1$ and $H_2$ is very strong compared to the complement and in particular compared to $H_3$, but that $H_1$ is only 1.8 times as likely as $H_2$ ($B_{13} = 1.777$).

To summarize the second application, our data demonstrated strong evidence for the prototype explanation and a lack of support for the same-gender bias explanation in explaining perceptions of discrimination against female applicants in the hiring process of computer specialist and nurses. The relative evidence for the prototype explanation depended on its exact formulation, but was at least 919 times stronger than for the same-gender bias explanation, and 91 times stronger than for the complement. However, further research is required to determine whether identification with a female victim has zero or a positive effect on perceived discrimination.

\subsection{Supplementary output}
When saving results from the \texttt{test\_hyp} function to an object it is possible to print additional supplementary output. This output is provided to support a deeper understanding of the method and the primary output outlined in the above subsections. We illustrate these two additional commands using the example in section 4.2. Calling \texttt{BF\_computation} prints the measures of relative fit ``$f$'' and complexity ``$c$'' in \eqref{fcE}-\eqref{fcI2} of the Bayes factor of each hypothesis against the unconstrained hypothesis. Thus, for the data and hypotheses of section 4.2 we get

\begin{verbatim}
result$BF_computation
##     c(E) c(I|E)     c  f(E) f(I|E)     f B(t,u) PP(t)
## H1 0.151  0.500 0.075 4.398  1.000 4.398 58.265 0.639
## H2    NA  0.020    NA    NA  0.650    NA 32.525 0.357
## H3 0.273  0.201 0.055 0.002  0.985 0.002  0.036 0.000
## Hc    NA  0.980    NA    NA  0.350    NA  0.357 0.004
\end{verbatim}

\noindent 
where \texttt{c(E)} is the prior density at the null value, \texttt{c(I|E)} the prior probability that the constraints hold, \texttt{c} the product of these two, and the columns labeled as \texttt{f(E)}, \texttt{f(I|E)}, and \texttt{f} have similar interpretations for the posterior quantities. \texttt{B(t,u)} is the Bayes factor of hypothesis $H_t$ against the unconstrained ($H_u$) and \texttt{PP(t)} is the posterior probability of hypothesis $H_t$. We rounded the output to three decimals for convenience. Cells with ``{\tt NA}'' indicate that a column is ``Not Available'' to a particular hypothesis. For example, because $H_2$ contains only inequality comparisons  it has a prior (and posterior) probability but no prior density evaluated at a null value. Hypothesis $H_1$ and $H_3$ contain both equality and inequality comparisons and thus has both prior and posterior densities and probabilities. The Bayes factor for $H_1$ and $H_3$ against $H_u$ can thus be calculated as $B_{1u}=\frac{4.398}{0.075}=58.265$ and $B_{3u}=\frac{0.002}{0.055}=0.036$ (see column \texttt{B(t,u)}). The posterior hypothesis probabilities are calculated using \eqref{PHP} by setting equal prior probabilities, i.e., $\text{Pr}(H_t|\textbf{y})=\frac{B_{tu}}{\sum_{t'} B_{t'u}}$, yielding, for example, $\text{Pr}(H_1|\textbf{y}) = \frac{58.265}{58.265+32.525+0.036+0.357} = 0.639$ (as indicated in column \texttt{PP(t)}).

If $\textbf{R}_I$ is not of full row rank, the posterior and prior that the inequality constraints hold are computed as the proportion of draws from unconstrained Student $t$ distributions. Under these circumstances there will be a, typically small, numerical Monte Carlo error. The 90\% credibility intervals of the numerical estimate of the Bayes factors of the hypotheses against the unconstrained hypothesis can be obtained by calling

\begin{verbatim}
result$BFu_CI
##    B(t,u) lb. (5%) ub. (95%)
## H1 58.265   58.169    58.360
## H2 32.525   32.152    32.910
## H3  0.036       NA        NA
## Hc  0.357    0.356     0.358
\end{verbatim}

where \texttt{B(t,u)} is the Bayes factor of hypothesis $t$ against the unconstrained ($u$), \texttt{lb. (5\%)} is the lower bound of the 90\% credibility interval estimate of the Bayes factor and \texttt{ub. (95\%)} is the upper bound. Credibility intervals are only printed when the computed Bayes factors have numerical errors. If the user finds the Monte Carlo error to be too large they can increase the number of draws from the Student $t$ distributions by adjusting the input value for the \texttt{mcrep} argument (default $10^6$ draws).

\section{Discussion}
The paper presented a new Bayes factor test for evaluating hypotheses on the relative effects in a linear regression model. The proposed testing procedure has several useful properties such as its flexibility to test multiple equality and/or order constrained hypotheses directly against each other, its intuitive interpretation as a measure of the relative evidence in the data between the hypotheses, and its fast computation. Moreover no prior information needs to be manually specified about the expected magnitude of the effects before observing the data. Instead, a default procedure is employed where a minimal fraction of the data is used for default prior specification and the remaining fraction is used for hypothesis testing. A consequence of this choice is that the statistical evidence cannot be updated using Bayes' theorem when observing new data. This is common in default Bayes factors \citep[e.g.,][]{OHagan:1997,Berger:2004}. Instead, the statistical evidence needs to be recomputed when new data are observed. This however is not a practical problem because of the fast computation of the default Bayes factor due to its analytic expression.

Furthermore the readily available {\tt lmhyp}-package can easily be used in combination with the popular {\tt lm}-package for linear regression analysis. The new method will allow researchers to perform default Bayesian exploratory analyses about the presence of a positive, negative or zero effect and to perform default Bayesian confirmatory analyses where specific relationships are expected between the regression effects which can be translated to equality and order constraints. The proposed test will therefore be a valuable contribution to the existing literature on Bayes factor tests \citep[e.g.,][]{Klugkist:2005,Rouder:2009,Klugkist:2010,Schoot:2011b,WetzelsWagenmakers2012PBR,Rouder:2012,Rouder:2015,Mulder:2012,Mulder:2014a,Gu:2014,Mulder:2016,BoingMessing:2017,Mulder:2018}, which are gradually winning ground as alternatives to classical significance tests in social and behavioral research. Due to this increasing literature, a thorough study about the qualitative and quantitative differences between these Bayes factors is called for. Another useful direction for further research would be to derive Bayesian (interval) estimates under the hypothesis that receives convincing evidence from the data.

\section*{Acknowledgements}
We would like to thank Stephen Wood for allowing us to use the empirical data in the first empirical application in this paper. These data are from the Service Delivery and Organisation (SDO) National Inpatient Staff Morale Study, which was funded by the National Institute for Health Research Service Delivery and Organisation (NIHR SDO) programme (Project Number 08/1604/142) and of which Stephen Wood was a co-investigator. The views and opinions expressed in the article are those of the authors and do not necessarily reflect those of the NIHR SDO programme or the Department of Health. The first author was supported by a NWO Vidi grant (452-17-006).

\appendix

\section{Proof of Lemma 1}
A derivation is given for the prior adjusted default Bayes factor for a hypothesis $H_1:\textbf{R}_E\bm\beta=\textbf{r}_E$, $\textbf{R}_I\bm\beta>\textbf{r}_I$ against an unconstrained hypothesis $H_u:\bm\beta\in\mathbb{R}^k$ in \eqref{fcE}-\eqref{fcI}. Based on the reparameterization $\bm\xi=\left[\begin{array}{c}\bm\xi_E\\ \bm\xi_I\end{array}\right]=\left[\begin{array}{c}\textbf{R}_E\\ \textbf{D}\end{array}\right]\bm\beta$ in \eqref{repara}, the hypothesis is equivalent to $H_1:\bm\xi_E=\textbf{r}_E$, $\tilde{\textbf{R}}_I\bm\xi_I>\tilde{\textbf{r}}_I$ against an unconstrained hypothesis $H_u:\bm\xi\in\mathbb{R}^k$. The marginal likelihood under the constrained hypothesis $H_1$ is defined as in the fractional Bayes factor \citep{OHagan:1995} with the exception that we integrate over an adjusted integration region \citep{Mulder:2014b,BoingMessing:2017}. This adjustment ensures that the implicit default prior is centered on the boundary of the constrained space. The marginal likelihood under $H_1$ is defined by
\begin{eqnarray}
p_1(\textbf{y},b) = \frac{\iint_{\tilde{\textbf{R}}_I\bm\xi_I>\tilde{\textbf{r}}_I} p(\textbf{y}|\bm\xi_E=\textbf{r}_E,\bm\xi_I,\sigma^2)\pi_u^N(\bm\xi_I,\sigma^2)d\bm\xi_I d\sigma^2}
{\iint_{\tilde{\textbf{R}}_I\bm\xi_I>\textbf{r}^*_I} p(\textbf{y}|\bm\xi_E=\hat{\bm\xi}_E,\bm\xi_I,\sigma^2)^b\pi_u^N(\bm\xi_I,\sigma^2)d\bm\xi_I d\sigma^2}.
\end{eqnarray}
As can be seen the adjustment implies that in the denominator the fraction of the likelihood is evaluated at $\hat{\bm\xi}_E$ instead of $\textbf{r}_E$ and the integration region equals
\[
\tilde{\textbf{R}}_I(\bm\xi_I-\hat{\bm\xi}_I+\bm\mu_I^0)>\tilde{\textbf{r}}_I
\Leftrightarrow
\tilde{\textbf{R}}_I\bm\xi_I>\tilde{\textbf{R}}_I\hat{\bm\xi}_I=\textbf{r}_I^*,
\]
because $\tilde{\textbf{R}}_I\bm\mu_I^0=\tilde{\textbf{r}}_I$, instead of $\tilde{\textbf{R}}_I\bm\xi_I>\tilde{\textbf{r}}_I$. Note that $\hat{\bm\xi}_I=\textbf{D}\hat{\bm\beta}$. This adjustment of the fractional Bayes factor ensures that the proposed default Bayes factor is computed using an implicit default prior that is centered on the boundary of the constrained space, following Jeffreys' heuristic argument \citep[see][for a more comprehensive motivation]{Mulder:2014b}. Furthermore this ensures that the complexity of an order-constrained hypothesis is properly incorporated in the Bayes factor \citep{Mulder:2014a}.
The marginal likelihood under $H_u$ is defined by
\begin{eqnarray}
p_u(\textbf{y},b) = \frac{\iiint p(\textbf{y}|\bm\xi_E,\bm\xi_I,\sigma^2)\pi_u^N(\bm\xi_E,\bm\xi_I,\sigma^2)d\bm\xi_E d\bm\xi_I d\sigma^2}
{\iiint p(\textbf{y}|\bm\xi_E,\bm\xi_I,\sigma^2)^b\pi_u^N(\bm\xi_E,\bm\xi_I,\sigma^2)d\bm\xi_E d\bm\xi_I d\sigma^2}.
\end{eqnarray}
The fraction $b$ will be set to $\frac{k+1}{n}$ because $k+1$ observations are needs to obtain a finite marginal likelihood when using a noninformative prior $\pi^N_u(\bm\xi_E,\bm\xi_I,\sigma^2)=\sigma^{-2}$ under $H_u$.

The default Bayes factor is then given by
\begin{eqnarray}
\nonumber B_{1u,b} &=& \frac{p_1(\textbf{y},b)}{p_u(\textbf{y},b)}
= \frac{\iint_{\tilde{\textbf{R}}_I\bm\xi_I>\tilde{\textbf{r}}_I} p(\textbf{y}|\bm\xi_E=\textbf{r}_E,\bm\xi_I,\sigma^2)\pi_u^N(\bm\xi_I,\sigma^2)d\bm\xi_I d\sigma^2}
{\iint_{\tilde{\textbf{R}}_I\bm\xi_I>\textbf{r}^*_I} p(\textbf{y}|\bm\xi_E=\hat{\bm\xi}_E,\bm\xi_I,\sigma^2)^b\pi_u^N(\bm\xi_I,\sigma^2)d\bm\xi_I d\sigma^2}\text{\Huge{/}}\\
\nonumber && \frac{\iiint p(\textbf{y}|\bm\xi_E,\bm\xi_I,\sigma^2)\pi_u^N(\bm\xi_E,\bm\xi_I,\sigma^2)d\bm\xi_E d\bm\xi_I d\sigma^2}
{\iiint p(\textbf{y}|\bm\xi_E,\bm\xi_I,\sigma^2)^b\pi_u^N(\bm\xi_E,\bm\xi_I,\sigma^2)d\bm\xi_E d\bm\xi_I d\sigma^2}\\
\nonumber &=& \iint_{\tilde{\textbf{R}}_I\bm\xi_I>\tilde{\textbf{r}}_I}\frac{ p(\textbf{y}|\bm\xi_E=\textbf{r}_E,\bm\xi_I,\sigma^2)\pi_u^N(\bm\xi_I,\sigma^2)}
{\iiint p(\textbf{y}|\bm\xi_E,\bm\xi_I,\sigma^2)\pi_u^N(\bm\xi_E,\bm\xi_I,\sigma^2)d\bm\xi_E d\bm\xi_I d\sigma^2}d\bm\xi_I d\sigma^2
\text{\Huge{/}}\\
\nonumber && \iint_{\tilde{\textbf{R}}_I\bm\xi_I>\textbf{r}^*_I}\frac
{ p(\textbf{y}|\bm\xi_E=\hat{\bm\xi}_E,\bm\xi_I,\sigma^2)^b\pi_u^N(\bm\xi_I,\sigma^2)}
{\iiint p(\textbf{y}|\bm\xi_E,\bm\xi_I,\sigma^2)^b\pi_u^N(\bm\xi_E,\bm\xi_I,\sigma^2)d\bm\xi_E d\bm\xi_I d\sigma^2}
d\bm\xi_I d\sigma^2\\
\nonumber &=& \iint_{\tilde{\textbf{R}}_I\bm\xi_I>\tilde{\textbf{r}}_I}\pi_u(\bm\xi_E=\textbf{r}_E,\bm\xi_I,\sigma^2|\textbf{y})
d\bm\xi_I d\sigma^2
\text{\Huge{/}}\\
\nonumber && \iint_{\tilde{\textbf{R}}_I\bm\xi_I>\textbf{r}^*_I}\pi_u(\bm\xi_E=\hat{\bm\xi}_E,\bm\xi_I,\sigma^2|\textbf{y}^b)
d\bm\xi_I d\sigma^2\\
\label{B1uApp}&=& \frac{\text{Pr}(\tilde{\textbf{R}}_I\bm\xi_I>\tilde{\textbf{r}}_I|\textbf{y},\bm\xi_E=\textbf{r}_E)}
{\text{Pr}(\tilde{\textbf{R}}_I\bm\xi_I>\textbf{r}^*_I|\textbf{y},\bm\xi_E=\hat{\bm\xi}_E)}\times 
\frac{\pi_u(\bm\xi_E=\textbf{r}_E|\textbf{y})}{\pi_u(\bm\xi_E=\hat{\bm\xi}_E|\textbf{y}^b)}.
\end{eqnarray}
Furthermore, using standard calculus it can be shown that the marginal posterior for $\bm\beta$ for a fraction $b$ of the data and a noninformative prior has a Student $t$ distribution
\begin{eqnarray*}
\pi_u(\bm\beta|\textbf{y}^b) &\propto &\int p(\textbf{y}|\bm\beta,\sigma^2)^b \pi_u^N(\bm\beta,\sigma^2) d\sigma^2\\
&\propto & t(\bm\beta;\hat{\bm\beta},s^2(nb-k)^{-1}(\textbf{X}'\textbf{X})^{-1},nb-k),
\end{eqnarray*}
and therefore, because $\bm\xi=\textbf{T}\bm\beta|\textbf{y}^b$, it holds that
\begin{eqnarray*}
\pi_u(\bm\xi|\textbf{y}^b) =
t(\bm\xi;\hat{\bm\xi},s^2(nb-k)^{-1}\textbf{T}(\textbf{X}'\textbf{X})^{-1}\textbf{T}',nb-k),
\end{eqnarray*}
where $\textbf{T}=\left[\begin{array}{c}\textbf{R}_E\\ \textbf{D}\end{array}\right]$, $\hat{\bm\xi}=(\hat{\bm\xi}_E',\hat{\bm\xi}_I')'$ with $\hat{\bm\xi}_E=\textbf{R}_E\hat{\bm\beta}$ and $\hat{\bm\xi}_I=\textbf{D}\hat{\bm\beta}$, and $t(\bm\xi;\textbf{m},\textbf{K},\nu)$ denotes a Student $t$ distribution for $\bm\xi$ with location parameters $\textbf{m}$, scale matrix $\textbf{K}$, and $\nu$ degrees of freedom.
Then it is well-known that the marginal distribution of $\bm\xi_E$ and the conditional distribution of $\bm\xi_I|\bm\xi_E$ also have Student $t$ distributions \citep[e.g.,][]{Press} given by
\begin{eqnarray}
\label{marg}\bm\xi_E|\textbf{y}^b & \sim &
t(\hat{\bm\xi}_E,s^2(nb-k)^{-1}\textbf{R}_E(\textbf{X}'\textbf{X})^{-1}\textbf{R}'_E,nb-k)\\
\label{cond}\bm\xi_I|\bm\xi_E,\textbf{y}^b & \sim &
t(\textbf{m}_{I|E},\textbf{K}_{I|E},\nu),
\end{eqnarray}
with
\begin{eqnarray*}
\textbf{m}_{I|E} &=& \textbf{D}\hat{\bm\beta}+s^{-2}(nb-k)\textbf{D}(\textbf{X}'\textbf{X})^{-1}\textbf{R}_E'(\textbf{R}_E(\textbf{X}\textbf{X})^{-1}\textbf{R}_E)^{-1}(\bm\xi_E-\hat{\bm\xi}_E)\\
\textbf{K}_{I|E} &=& \frac{nb-k + s^{-2}(nb-k)(\bm\xi_E-\hat{\bm\xi}_E)'(\textbf{R}_E'(\textbf{X}'\textbf{X})^{-1}\textbf{R}_E)^{-1}(\bm\xi_E-\hat{\bm\xi}_E)}{nb-k+q^E}\\
&&(s^2(nb-k)^{-1}\textbf{D}(\textbf{X}'\textbf{X})^{-1}\textbf{D}'-\\
&& s^2(nb-k)^{-1}
\textbf{D}(\textbf{X}'\textbf{X})^{-1}\textbf{R}_E'(\textbf{R}_E(\textbf{X}'\textbf{X})^{-1}\textbf{R}_E')^{-1}\textbf{R}_E(\textbf{X}'\textbf{X})^{-1}\textbf{D}).
\end{eqnarray*}

Thus, when plugging in $b=1$ and $\bm\xi_E=\textbf{r}_E$ in \eqref{marg} and \eqref{cond}, and then in \eqref{B1uApp}, gives the numerators in \eqref{fcE} and \eqref{fcI}, and plugging in $b=\frac{k+1}{n}$ and $\bm\xi_E=\hat{\bm\xi}_E$ in \eqref{marg} and \eqref{cond}, and then in \eqref{B1uApp}, gives the denominators in \eqref{fcE} and \eqref{fcI}, which completes the proof.

\section{Example analysis for Figure 1}
\begin{verbatim}
# consider a regression model with two predictors:
# y_i = beta_0 + beta_1 * x1_i + beta_2 * x2_i + error
library(lmhyp)
n <- 20 #sample size
X <- mvtnorm::rmvnorm(n,sigma=diag(3))
# For this example we transform X to get exact independent
# predictor variables and errors.
X <- X - rep(1,n)%*%t(apply(X,2,mean))
X <- X%*%solve(chol(t(X)%*%X))*sqrt(n)
errors <- X[,3] #a population variance of 1
X <- X[,1:2]
beta <- c(.7,.03) #data generating regression effects
y <- 1 + X%*%beta + errors
df1 <- data.frame(y=y,x1=X[,1],x2=X[,2])
fit1 <- lm(y~x1+x2,df1)
test1 <- test_hyp(fit1,"x1=x2=0;(x1,x2)>0;x1>x2=0")
test1 #get posterior probabilities
test1$BF_matrix #get Bayes factors
test1$	BF_computation #get details on the computations
test1$BFu_CI #get 90% credibility intervals, if applicable
\end{verbatim}

\bibliographystyle{apacite}
\bibliography{refs_mulder}

\end{document}